\documentclass [twocolumn]{aastex701} %,linenumbers,trackchanges

\usepackage{amssymb, amsmath, mathtools}
\usepackage{enumerate}
\usepackage{graphicx}

\usepackage{latexsym}
\usepackage[T1]{fontenc}
\usepackage{tabularx}
\usepackage{multirow}
\usepackage{colortbl}
\usepackage[version=4]{mhchem} % Molecular species! e.g. \ce{H2O}
\usepackage{xspace}
\usepackage{color}
\usepackage{graphicx}
\usepackage{hyperref} % For hyperlinks in the PDF
\usepackage[nameinlink]{cleveref}

\newcommand\degree{\degr}
\newcommand\degrees\degree

\newcommand\jwst{\em JWST}

\newcommand\eureka{\texttt{Eureka!}\xspace}

\newcommand{\poseidon}{\texttt{POSEIDON}\xspace}

\DeclareSymbolFont{UPM}{U}{eur}{m}{n}
\DeclareMathSymbol{\umu}{0}{UPM}{"16}
\let\oldumu=\umu
\renewcommand\umu{\ifmmode\oldumu\else\math{\oldumu}\fi}

\newcommand\microns \micron

\let\oldsim=\sim
\renewcommand\sim{\ifmmode\oldsim\else\math{\oldsim}\fi}
\let\oldpm=\pm
\renewcommand\pm{\ifmmode\oldpm\else\math{\oldpm}\fi}
\newcommand\by{\ifmmode\times\else\math{\times}\fi}

\newbox{\wdbox}
\renewcommand\c{\setbox\wdbox=\hbox{,}\hspace{\wd\wdbox}}
\renewcommand\i{\setbox\wdbox=\hbox{i}\hspace{\wd\wdbox}}
\newcommand\n{\hspace{0.5em}}

\newcount\timect
\newcount\hourct
\newcount\minct
\newcommand\now{\timect=\time \divide\timect by 60
         \hourct=\timect \multiply\hourct by 60
         \minct=\time \advance\minct by -\hourct
         \number\timect:\ifnum \minct < 10 0\fi\number\minct}

\catcode`@=11

\newcommand\comment[1]{}
\newcommand\commenton{\catcode`\%=14}
\newcommand\commentoff{\catcode`\%=12}

\renewcommand\math[1]{$#1$}
\newcommand\mathshifton{\catcode`\$=3}
\newcommand\mathshiftoff{\catcode`\$=12}

\comment{the backslash is necessary}

\comment{alignment tab}

\let\atab=&
\newcommand\atabon{\catcode`\&=4}
\newcommand\ataboff{\catcode`\&=12}

\let\oldmsp=\sp
\let\oldmsb=\sb
\def\sp#1{\ifmmode
           \oldmsp{#1}%
         \else\strut\raise.85ex\hbox{\scriptsize #1}\fi}
\def\sb#1{\ifmmode
           \oldmsb{#1}%
         \else\strut\raise-.54ex\hbox{\scriptsize #1}\fi}
\newbox\@sp
\newbox\@sb
\def\sbp#1#2{\ifmmode%
           \oldmsb{#1}\oldmsp{#2}%
         \else
           \setbox\@sb=\hbox{\sb{#1}}%
           \setbox\@sp=\hbox{\sp{#2}}%
           \rlap{\copy\@sb}\copy\@sp
           \ifdim \wd\@sb >\wd\@sp
             \hskip -\wd\@sp \hskip \wd\@sb
           \fi
        \fi}
\def\msp#1{\ifmmode
           \oldmsp{#1}
         \else \math{\oldmsp{#1}}\fi}
\def\msb#1{\ifmmode
           \oldmsb{#1}
         \else \math{\oldmsb{#1}}\fi}
\def\supon{\catcode`\^=7}
\def\supoff{\catcode`\^=12}
\def\subon{\catcode`\_=8}
\def\suboff{\catcode`\_=12}
\def\supsubon{\supon \subon}
\def\supsuboff{\supoff \suboff}

\newcommand\actcharon{\catcode`\~=13}
\newcommand\actcharoff{\catcode`\~=12}

\newcommand\paramon{\catcode`\#=6}
\newcommand\paramoff{\catcode`\#=12}

\comment{And now to turn us totally on and off...}

\newcommand\reservedcharson{\commenton \mathshifton \atabon \supsubon \actcharon
	\paramon}

\newcommand\reservedcharsoff{\commentoff \mathshiftoff \ataboff
	\supsuboff \actcharoff \paramoff}

\catcode`@=12
\reservedcharsoff

\reservedcharson

\comment{ Must have ONLY ONE of these... trust these macros, they work

}

\newenvironment{packed_enum}{
\begin{enumerate}
   \setlength{\itemsep}{1pt}
   \setlength{\parskip}{0pt}
   \setlength{\parsep}{0pt}
}{\end{enumerate}}

\newcommand{\squishlist}{
 \begin{list}{$\bullet$}
  { \setlength{\itemsep}{0pt}
     \setlength{\parsep}{0pt}
     \setlength{\topsep}{0pt}
     \setlength{\partopsep}{0pt}
     \setlength{\leftmargin}{2.0em}
     \setlength{\labelwidth}{1.5em}
     \setlength{\labelsep}{0.5em} } }

\newcommand{\squishlisttwo}{
 \begin{list}{$\bullet$}
  { \setlength{\itemsep}{1pt}
     \setlength{\parsep}{3pt}
     \setlength{\topsep}{3pt}
     \setlength{\partopsep}{0pt}
     \setlength{\leftmargin}{2.0em}
     \setlength{\labelwidth}{1.5em}
     \setlength{\labelsep}{0.5em} } }

\newcommand{\squishend}{
  \end{list}  }

\reservedcharson
\actcharon

\newcommand{\figsetcapnum}{}
\newcommand{\figsetcaptitle}{}

\renewcommand{\figsetgrpstart}{\begin{figure*}[!h]\renewcommand{\figurename}{Fig.}\renewcommand{\thefigure}{Set \figsetcapnum~$-$ \figsetcaptitle}
}
\renewcommand{\figsetgrpend}{\end{figure*}}
\graphicspath{{./}{figs/}}

\newcommand{\planetname}{K2-18b\xspace}

\shorttitle{\planetname Does Not Meet The Standards of Evidence For Life}
\shortauthors{Stevenson et al.}
\newcommand{\APL}{Johns Hopkins APL, 11100 Johns Hopkins Rd, Laurel, MD 20723, USA}
\newcommand{\Goddard}{NASA Goddard Space Flight Center, 8800 Greenbelt Road, Greenbelt, MD 20771, USA}
\newcommand{\CHAMPs}{Consortium on Habitability and Atmospheres of M-dwarf Planets (CHAMPs), Laurel, MD, USA}

\begin{document}

% \title{No Evidence for Life on \planetname}
\title{\planetname Does Not Meet The Standards of Evidence For Life}

\correspondingauthor{Kevin Stevenson}
\email{Kevin.Stevenson@jhuapl.edu}

\author[0000-0002-7352-7941]{Kevin B. Stevenson}
\affiliation{\APL}
\affil{\CHAMPs}
\email{Kevin.Stevenson@jhuapl.edu}

\author[0000-0002-0746-1980]{Jacob Lustig-Yaeger}
\affiliation{\APL}
\affil{\CHAMPs}
\email{Jacob.Lustig-Yaeger@jhuapl.edu}

\author[0000-0002-2739-1465]{E. M. May}
\affiliation{\APL}
\affil{\CHAMPs}
\email{Erin.May@jhuapl.edu}

\author[0000-0002-5893-2471]{Ravi K. Kopparapu}
\affiliation{\Goddard}
\affil{\CHAMPs}
\email{ravikumar.kopparapu@nasa.gov}

\author[0000-0002-5967-9631]{Thomas J. Fauchez}
\affiliation{\Goddard}
\affiliation{Integrated Space Science and Technology Institute, Department of Physics, American University, Washington DC}
\affil{\CHAMPs}
\email{thomas.j.fauchez@nasa.gov}

\author[0000-0003-4346-2611]{Jacob Haqq-Misra}
\affiliation{Blue Marble Space Institute of Science, Seattle, WA, USA}
\affil{\CHAMPs}
\email{jacob@bmsis.org}

\author[0000-0002-9521-9798]{Mary Anne Limbach}
\affiliation{Department of Astronomy, University of Michigan, Ann Arbor, MI 48109, USA}
\affil{\CHAMPs}
\email{mlimbach@umich.edu}

\author[0000-0002-2949-2163]{Edward W. Schwieterman}
\affiliation{Department of Earth and Planetary Sciences, University of California, Riverside, CA, USA}
\affiliation{Blue Marble Space Institute of Science, Seattle, WA, USA}
\affil{\CHAMPs}
\email{eschwiet@ucr.edu}

\author[0000-0001-7393-2368]{Kristin S. Sotzen}
\affiliation{\APL}
\affil{\CHAMPs}
\email{Kristin.Sotzen@jhuapl.edu}

\author[0000-0002-8163-4608]{Shang-Min Tsai}
\affiliation{Institute of Astronomy \& Astrophysics, Academia Sinica, Taipei 10617, Taiwan}
\affil{\CHAMPs}
\email{smtsai@asiaa.sinica.edu.tw}

\begin{abstract}

% While Jeff Goldblum once assured us that ``life finds a way," K2-18b offers a different lesson: instrument systematics find a way.

\planetname, a temperate sub-Neptune, has garnered significant attention due to claims of possible biosignatures in its atmosphere.  Low-confidence detections of dimethyl sulfide (DMS) and/or dimethyl disulfide (DMDS) have sparked considerable debate, primarily around arguments that their absorption features are not uniquely identifiable. Here, we consider all five questions from the astrobiology standards of evidence framework, starting with: Have we detected an authentic signal? To answer this, we analyzed publicly-available {\jwst} observations of \planetname using independent data reduction and spectral retrieval methodologies.  Our comprehensive set of reductions demonstrates that the MIRI transit spectrum is highly susceptible to unresolved instrumental systematics.  Applying different wavelength binning schemes yields a potpourri of planet spectra that then lead to a wide assortment of atmospheric interpretations.  Consequently, we offer recommendations to help minimize this previously-underappreciated instrument systematic in future MIRI reductions of any exoplanet. While the MIRI binning scheme adopted by \citet{Madhusudhan2025} favors the presence of DMS/DMDS in \planetname, we find that 87.5\% of retrievals using our preferred MIRI binning scheme do not. When considering the full, 0.7--12 {\micron} transit spectrum, we confirm the detection of \ce{CH4} and favor \ce{CO2}, and find the presence of DMS and \ce{C2H4} to be interchangeable. Moreover, we find that the tentative presence of large features in the MIRI transit spectrum is in tension with the more robust, yet smaller, features observed in the near IR.  We conclude that red noise --- rather than an astrophysical signal --- plagues the mid-IR data and there is, as yet, no statistically significant evidence for biosignatures in the atmosphere of \planetname.

\end{abstract}

% \keywords{methods: data analysis{$:$} 
% planets and satellites: atmospheres}

\section{Introduction}
\label{sec:intro}

\subsection{Astrobiology Standards of Evidence}
\label{sec:intro:evidence}

High-precision spectroscopy from observatories such as the James Webb Space Telescope ({\jwst}) has facilitated the detection of molecular signatures in exoplanet atmospheres, leading to intriguing, but also potentially premature, claims about habitability and life. Such claims are not unique to the exoplanet field: prior assertions of detecting extraterrestrial life have either been contradicted or remain under debate. Examples include the Viking lander's search for microorganisms on the Martian surface \citep[e.g.,][]{levin2016case}, potential fossilized microbial life in Martian meteorite ALH 84001 \citep[e.g.,][]{mcKay1996}, and the apparent detection of phosphine gas in Venus’s atmosphere \citep[e.g.,][]{greaves2021}.

In each of these cases, community consensus challenged the original claims without confirming the existence of life beyond Earth. This is unsurprising for a simple reason: the detection of extraterrestrial life is unlikely to be rapid or straightforward \citep[see, e.g.,][]{dick2018astrobiology}. Such claims require rigorous scientific scrutiny, including the collection of additional data, multiple independent analyses, and interdisciplinary collaboration. A systematic, structured framework for evaluating these claims would greatly benefit the exoplanet field, helping the community clearly communicate results to all stakeholders, ranging from scientists to the broader public.

To that end, a comprehensive, interdisciplinary effort by the astrobiology community produced two key documents: (1) a call for developing a confidence scale for biosignature detection \citep{green2021}, and (2) a community-organized report proposing a generalized framework for evaluating biosignature claims \citep{meadows2022}. The latter, co-authored by experts in astrophysics, planetary science, chemistry, biology, and data analysis, presents five guiding questions that should be iteratively addressed before inferring life from remote observations. These questions form the basis for a community consensus on biosignature detection standards. We restate them here for emphasis and convenience:

\begin{itemize}
    \item {\bf Q1: Have you detected an authentic signal?} Have you authenticated your signal, and is it statistically significant? Have you ruled out artifacts from the measurement, pre-processing and/or analysis process that might mimic a real signal?
    \item {\bf Q2: Have you adequately identified the signal?} Have you adequately ruled out other potential sources for this signal? For example, have you ruled out contamination in the environment, or other real phenomena that could produce a similar signal?
    \item {\bf Q3: Are there abiotic sources for your detection?} Is it likely that there is a current or past environmental process, other than life, that could be producing this signal? Have you ruled out these potential false positives for the biosignature?
    \item {\bf Q4: Is it likely that life would produce this expression in this environment?} Given what we know about the likely environment that an organism is operating in, or would have operated in, does it make physical and chemical sense that life would produce this potential biosignature?
    \item {\bf Q5: Are there independent lines of evidence to support a biological (or non-biological) explanation?} Are there other measurements that provide additional evidence, or allow you to predict and execute follow-on experiments, that will help discriminate between the life or non-life hypotheses?
\end{itemize}

This structured approach builds upon the earlier guidance by \citet{green2021} and aims to reduce ambiguity and prevent premature conclusions. It is not intended to be restrictive, but to encourage transparency, interdisciplinary engagement, and reproducibility as the field shifts from theoretical to observational biosignature science. 
Importantly, the framework promotes an iterative, rather than linear, approach. A proliferation of false positives would not only stall progress by forcing stricter detection thresholds, but could also erode public trust in astrobiology and the scientific process as a whole. Upholding rigorous, transparent standards is therefore essential to ensure that any reported evidence for life beyond Earth is both credible and enduring.

In this context, recent claims \citep[][hereafter M23 and M25]{Madhusudhan2023,Madhusudhan2025} of dimethyl sulfide (DMS) and/or dimethyl disulfide (DMDS) --- volatile sulfur-bearing molecules associated with biological activity on Earth --- in the atmosphere of the sub-Neptune \planetname merit careful evaluation. 
As a proof of concept, we adopt the astrobiology standards of evidence framework to assess these claims. In particular, we evaluate whether the claims satisfactorily address {\bf Q1} (signal authenticity) and {\bf Q2} (signal attribution). We also discuss the utility of adopting this framework for future biosignature claims in exoplanet data.

\subsection{SETI Standards of Evidence}
\label{sec:intro:seti}

It is worth noting that the search for extraterrestrial intelligence (SETI) community, through decades of effort to detect radio or optical signals from alien civilizations, has also developed evidence-based protocols for handling candidate detections. The first such protocols were adopted in 1989 by the SETI Permanent Committee of the International Academy of Astronautics (IAA)\footnote{\href{https://iaaseti.org/en/declaration-principles-concerning-activities-following-detection/}{Declaration of Principles Concerning Activities Following the Detection of Extraterrestrial Intelligence (1989)}} and revised in 2010\footnote{\href{http://resources.iaaseti.org/protocols_rev2010.pdf}{Declaration of Principles Concerning the Conduct of the Search for Extraterrestrial Intelligence (2010)}}. 
The most recent post-detection protocols include eight principles, many of which align closely with the five astrobiology questions in \autoref{sec:intro:evidence}. For instance, the protocols recommend validating candidate signals in coordination with other SETI groups before announcing a detection. If the signal remains viable, the discovering scientists may issue a public announcement while continuing to monitor the signal and report progress at scientific meetings.

Although radio SETI and astrobiology are separate and distinct disciplines, the astrobiology standards of evidence guidelines build on several of the principles originally articulated in the SETI protocols. It is understood that these (and other) frameworks are not expected to be perfect and may require refinement. Furthermore, their adoption may also be challenged by differing views within the scientific community \citep{ind2022}. Nonetheless, they should serve as guiding questions --- rather than inviolable rules --- when evaluating claims of life beyond Earth.

\subsection{\planetname and the Debate on Life}
\label{sec:intro:k2-18b}

The sub-Neptune exoplanet, \planetname, has emerged as a high-profile target for atmospheric characterization with {\jwst}, following initial detections of \ce{CH4} and \ce{CO2} in a hydrogen-rich atmosphere \citep{Madhusudhan2023} and tentative claims of the biosignature gas, DMS. These observations can be interpreted as being consistent with a Hycean world, where a liquid water ocean resides beneath a hydrogen envelope.  Subsequently, M25 reported a mid-IR transit spectrum and retrieval analysis suggesting the presence of DMS and/or DMDS at $\sim3\sigma$ confidence, with mixing ratios on the order of tens of ppmv --- several orders of magnitude higher than DMS levels in Earth’s atmosphere \citep{hulswar2022}.

Following M23, \citet{tsai2024} used a self-consistent climate-photochemical model to assess the distribution and detectability of sulfur and methane-based gases on Hycean worlds like \planetname.  They found that a biogenic sulfur flux $\sim20\times$ higher than modern Earth's would be required for DMS to reach detectable levels (i.e., $\gtrsim 1$ppm) in the atmosphere of \planetname. They also predicted that such high concentrations should be accompanied by significant amounts of ethane, \ce{C2H6}, which is a photochemical byproduct of DMS photolysis. Indeed, \cite{Luque2025} propose \ce{C2H6} as an alternative explanation for the DMS/DMDS feature, though M25 did not report its detection despite its prominent mid-IR features.

The claims made by M23 and M25 have sparked both interest and debate within the exoplanet community. \citet{Schmidt2025} conducted an independent analysis of publicly available NIRISS/SOSS and NIRSpec/G395H transmission spectra, confirming \ce{CH4} but finding no statistically significant evidence for \ce{CO2} or DMS. They suggested that \planetname's composition is consistent with an oxygen-poor sub-Neptune, without requiring a liquid-water surface or life. Following M25’s publication, multiple studies revisited the MIRI/LRS data using independent retrieval frameworks and expanded molecular libraries. \citet{Taylor2025} and \citet{Luque2025} both cautioned that there is no statistical evidence for DMS or DMDS. The features lie near the noise threshold and are not uniquely attributable, as several other sulfur- or carbon-bearing species fit the data comparably well. \citet{Welbanks2025}, using an agnostic retrieval approach, found that the current data do not robustly constrain biosignature gases and are also consistent with various non-biogenic molecules. In contrast, \citet{PicaCiamarra2025} conducted a broad search for trace gases using a library of 650 molecules and found moderate evidence ($\ge2.7\sigma$) for numerous molecules --- including diethyl sulfide and methyl acrylonitrile --- when no offset was applied between NIRISS and NIRSpec. 

Using four new NIRSpec transit observations, \citet{Hu2025} robustly detect both \ce{CH4} and \ce{CO2}, and inferred that \planetname has a water-rich interior. They reported only marginal evidence for DMS and, through photochemical modeling, identified potential abiotic pathways for its formation. As examples of its abiotic prominence, DMS was recently found in comet 67P/Churyumov–Gerasimenko \citep{Hanni2024} and the interstellar medium \citep{SanzNovo2025}.

\citet{Seager2025} used their three key criteria for definitive exoplanet findings to rebuff the claims of M23.  Their questions (which relate to detection, attribution, and interpretation) are conceptually similar to the five standards of evidence questions in \autoref{sec:intro:evidence}, but are oriented toward exoplanet observations.  They conclude that \planetname fails all three criteria and emphasize the need for further exploration of abiotic production scenarios.  

Taken together, these findings highlight that while \planetname remains a compelling target for follow-up, current claims of biogenic gases are not yet robust.

In this paper, we re-reduce and reanalyze the publicly available MIRI, NIRSpec, and NIRISS data for \planetname (\autoref{sec:data}) and place new constraints on its atmospheric composition (\autoref{sec:modeling}). We then evaluate how the findings relate to the astrobiology standards of evidence framework (\autoref{sec:discussion}) and draw conclusions about the reliability of the MIRI spectra, the source of its spectral features, and the strength of evidence for biosignature gases (\autoref{sec:conclusion}).

\begin{figure*}[t]
    \centering
    \includegraphics[width=0.9\linewidth]{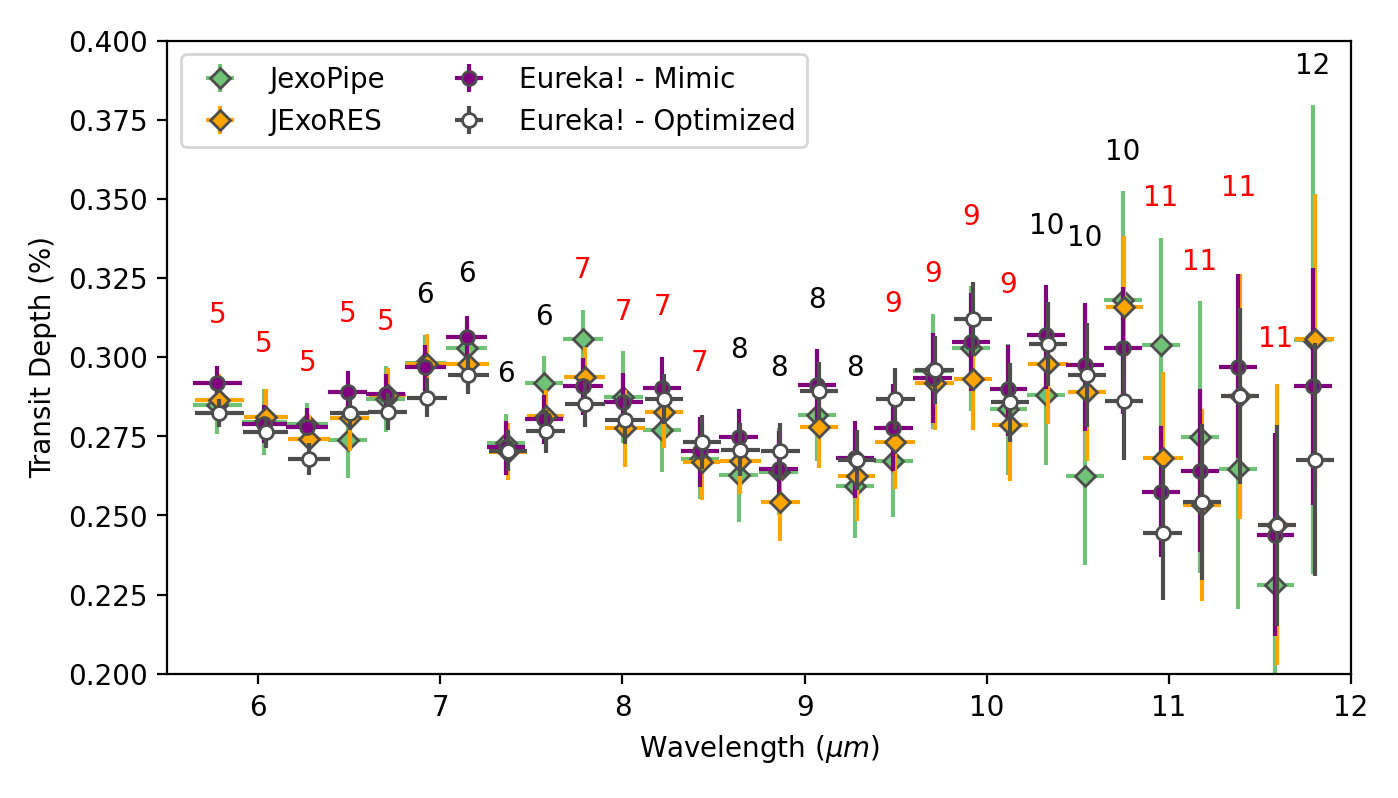}
    \caption{A comparison between \planetname MIRI transmission spectra.  The JexoPipe and JExoRES spectra are from M25. {\eureka}'s Mimic spectrum adopts the same settings described by M25. {\eureka}'s Optimized spectrum uses settings that minimize the white light curve MAD value.  When adopting the same wavelength binning scheme, the spectra from both {\eureka} reductions are comparable (i.e., within $1\sigma$) to those from M25.  The numerical values indicate the number of pixel rows used in a given spectroscopic channel.  As discussed in \autoref{sec:data:tests}, channels with red numbers have the least reliable transit depths.
    }
    \label{fig:MIRI_Compare}
\end{figure*}

\section{Data Reduction} 
\label{sec:data}

\subsection{MIRI}
\label{sec:data:MIRI}

We used the \eureka\ pipeline \citep[v1.2.1;][]{Bell2022} to reduce the {\jwst} MIRI/LRS data from GO program 2722 (PI: Madhusudhan). Standard data reduction techniques, previously applied to other MIRI datasets \citep[e.g.,][]{Malsky2025}, were used. We ran the \texttt{jwst} pipeline version 1.18.0 with CRDS reference context (i.e., {\em pmap}) 1364.

We first attempted to reproduce (or mimic) the spectra published by M25 by adopting the same reduction settings described therein. For any parameter not specified by M25, we selected a reasonable estimate. The time-series data were binned into 29 spectroscopic light curves using the same wavelength bins as M25. For the light curve fits, we trimmed the first 250 integrations and modeled the time-dependent systematics using an exponential plus linear component, as in M25. As shown in \Cref{fig:MIRI_Compare}, we were able to reproduce the spectra published by M25 when using an identical binning scheme.

\begin{figure*}[t]
    \centering
    \includegraphics[width=\linewidth]{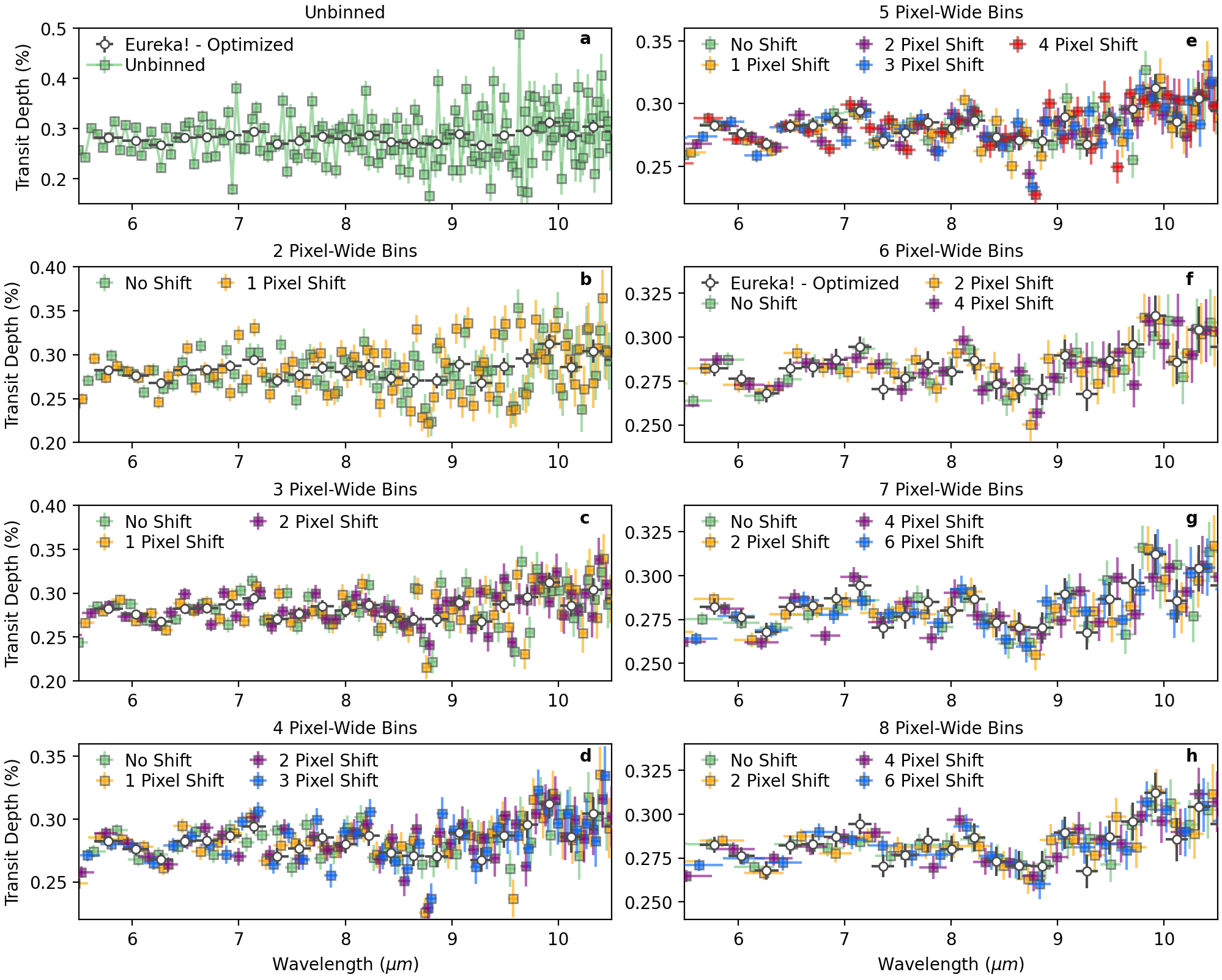}
    \caption{Measured MIRI/LRS transmission spectra of \planetname using a range of fixed channel sizes (1 -- 8 pixel-wide bins) and starting wavelengths (i.e., pixel shifts). The unbinned spectrum (panel {\bf a}) exhibits significant scatter between adjacent channels, with a mean absolute difference of 663 ppm. For reference, the typical feature size in \Cref{fig:MIRI_Compare} is $\sim250$ ppm.  The scatter in the measured spectrum is seen to decrease with larger channel sizes and an even number of pixel rows (panels {\bf b} -- {\bf h}).
    }
    \label{fig:MIRI_bins}
\end{figure*}

Next, we used \eureka's optimizer \citep{Ashtari2025} to perform a parametric sweep of 11 settings in Stage 3. The optimizer identified values that minimized the median absolute difference (MAD) of the white light curve. In this case, the white light curve MAD value improved from 415 ppm to 405 ppm. For these fits, we trimmed the first 1,200 integrations --- where the data exhibit an exponential decrease in flux --- and modeled the remaining integrations with a linear component. Overall, our ``Optimized'' spectrum is broadly consistent with that of M25 (\Cref{fig:MIRI_Compare}); however, we note important differences at specific wavelengths. M25’s leave-one-out analysis identified two regions (6.8 -- 7.3 {\microns} and 8.5 - 9.0 {\microns}; see their Figure 6) that are better predicted by a model including DMS/DMDS. In both regions, our ``Optimized'' spectrum exhibits smaller deviations from the mean transit depth (i.e., a flatter spectrum), thereby reducing the evidence for DMS/DMDS (\autoref{sec:modeling}). Trimming only 250 integrations and fitting an exponential ramp to the optimized reduction yields slightly different fluctuations in transit depth, but all are consistent at the $1\sigma$ level. Hereafter, we use the optimized reduction with a linear model component for all subsequent tests and analyses. We also adopt the best-fit orbital parameters from the NIRSpec/G395H reduction listed in \Cref{tab:fit_parameters}.

\subsubsection{Testing Different Binning Schemes} 
\label{sec:data:tests} 

Our overarching goal for this test is to assess the impact of different binning schemes on the measured \planetname\ transit spectrum. \citet{Bell2024} reported excess scatter in their unbinned L168-9b transit spectrum (see their Extended Data Fig. 1) and demonstrated that binning in wavelength space reduces this scatter. We conduct a similar analysis using fixed bin widths spanning 1 -- 8 pixel rows. Hereafter, we refer to each bin as a ``channel.'' As shown in \Cref{fig:MIRI_Compare}, M25 used variable bin widths ranging from 5 to 12 pixel rows per channel.

\begin{figure}[tbh]
    \centering
    \includegraphics[width=\linewidth]{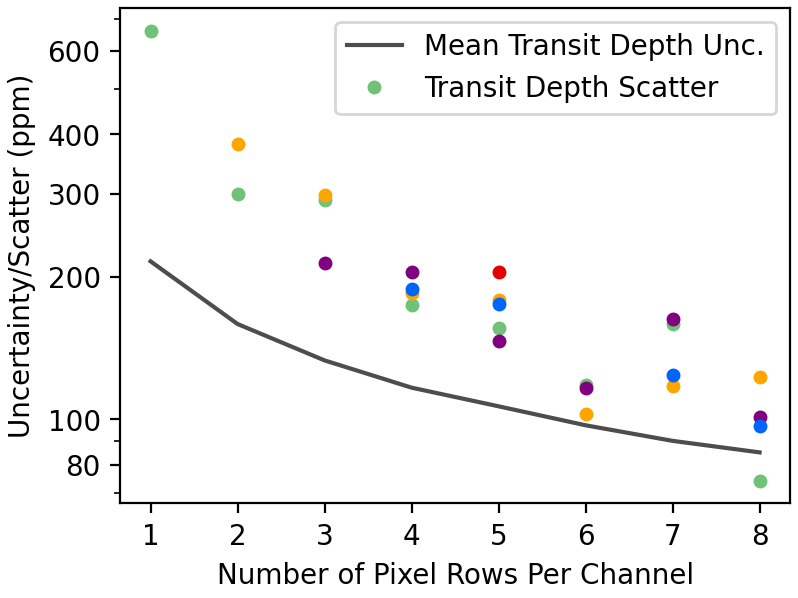}
    \caption{Observed transit depth scatter (colored circles) compared to the mean transit depth uncertainty (black curve). We quantify the scatter by computing the mean absolute difference (MAD) between adjacent channels.  The colors represent different pixel shifts and correspond to those shown in \Cref{fig:MIRI_bins}.
    For a low SNR spectrum, like that of \planetname, the scatter should not significantly exceed the mean transit depth uncertainty.  Furthermore, the scatter should be relatively consistent between pixel shifts.  For this test, we achieve the most robust transit depths with six or eight pixel rows per channel; however, we caution against simply adopting these binning schemes for other datasets without first conducting similar tests.
    }
    \label{fig:scatter}
\end{figure}

We begin by examining the unbinned transit spectrum shown in panel {\bf a} of \Cref{fig:MIRI_bins} and note significant offsets in the measured depths between adjacent channels. The mean difference between odd- and even-indexed channels is 441 ppm for data spanning 5 -- 10.5 {\microns}. The mean absolute difference between adjacent channels over the same range is even larger: 663 ppm. We refer to this quantity as the ``transit depth scatter'' in \Cref{fig:scatter}, and it is $3.1\times$ greater than the mean transit depth uncertainty.

Next, we adopt 2-pixel-wide bins (\Cref{fig:MIRI_bins}, panel {\bf b}) and observe a reduction in scatter (299 -- 382 ppm). However, we find a significant difference in the measured transit depths when the starting wavelength is shifted by one pixel (see green vs. orange squares). This suggests that assuming a simple odd/even pixel row offset is overly simplistic. These results motivate the exploration of larger channel sizes that, when shifted in wavelength, do not exhibit statistically significant differences in the measured depths.

To test the reliability of the measured spectra, we examine larger channel sizes (3 -- 8 pixel-wide bins; see panels {\bf c}–{\bf h} of \Cref{fig:MIRI_bins}) and a range of pixel shifts (0 -- 6 pixels). \Cref{fig:scatter} plots the transit depth scatter as a function of channel size for each pixel shift. For a flat transit spectrum, the scatter is expected to match the mean transit depth uncertainty (black curve). Excess scatter could arise from either astrophysical absorption features or instrumental systematics. In the former case, the scatter should remain relatively consistent across pixel shifts.

Inspection of \Cref{fig:scatter} reveals that channels composed of an odd number of pixel rows yield systematically larger transit depth scatter. This is consistent with our findings from the unbinned transit spectrum (panel {\bf a} of \Cref{fig:MIRI_bins}) and with the binning test of M25 (see teal points in their Figure 9). The implication here is that using an odd number of pixel rows per channel can introduce spurious signals into the measured transit spectrum.

\Cref{fig:scatter} also shows that bin sizes of five or fewer pixel rows can similarly introduce spurious signals due to instrumental systematics. We conclude that, for this MIRI/LRS dataset, an even number of pixel rows per channel --- at minimum six --- is necessary to mitigate these effects. For the remainder of this paper, we adopt a bin size of eight pixel rows per channel.

We note that, across the eight pixel shifts used for atmospheric modeling (\autoref{sec:modeling}), the measured scatter still varies substantially (MAD = 74 -- 153 ppm), indicating that further binning may be required. We recommend that future MIRI analyses explore a wider range of binning schemes and aim to achieve more consistent MAD values across pixel shifts. 

% Discuss M25's binning test
M25 reported in their binning test that their transit spectra are generally consistent across different bin widths and that atmospheric retrievals using these spectra yield similar DMS/DMDS detections when applying their canonical model (i.e., including only \ce{CH4}, \ce{CO2}, DMS, and DMDS). Such consistency is to be expected when doubling or quadrupling bin widths without varying the pixel shift.

\subsection{NIRSpec} 
\label{sec:data:nirspec} 
We use \eureka v1.2 \citep{Bell2022} to reduce the {\jwst} NIRSpec/G395H data from GO program 2722 (PI: Madhusudhan). This version of \eureka supports the \texttt{jwst} pipeline version 1.18.0. We use CRDS reference context (\textit{pmap}) 1364.

NIRSpec/G395H features a curved spectral trace that spans two detectors with a gap in between. We treat the two detectors, NRS1 and NRS2, as independent observations. For both detectors --- and similar to the MIRI reduction --- we optimize parameters in Stages 1, 3, and 4 of the \eureka pipeline to minimize the MAD values of the white light curve. 
The latest version of the \texttt{jwst} pipeline includes faster ramp fitting, allowing us to efficiently optimize parameters related to group-level background subtraction (GLBS). GLBS is an important step for the NIRSpec instrument, as it improves the precision of ramp fitting and, in turn, the precision of light curves. We otherwise run all standard steps in the \texttt{jwst} Stage 1 and Stage 2 pipeline for time series observations, while \eureka Stage 3 extracts the time-series stellar spectra.

\begin{deluxetable*}{llccccc}[t]
\tablewidth{0pt}
\tablecaption{K2-18 parameters from our white light curve fits. \label{tab:fit_parameters}}
\tablehead{
Parameter                   & Unit      & \multicolumn{2}{c}{NIRISS/SOSS} & \multicolumn{2}{c}{NIRSpec/G395H} & MIRI/LRS \\
& & Order 1 & Order 2 & NRS1 & NRS2 &  
}
\startdata
Radius Ratio, R$_p$/R$_s$   & \nodata   & 0.05415(13) & 0.05494(46) & 0.05403(10) &  0.05374(11)  &  0.05285(13) \\
Transit Midpoint, $t_0$     & MJD       & \multicolumn{2}{c}{60096.72931(7)} &  \multicolumn{2}{c}{59964.96948(3)}    &  60246.12873(8) \\
Inclination, $i$            & degrees   & \multicolumn{2}{c}{\nodata} & \multicolumn{2}{c}{89.604(11)} & \nodata \\
Semi-Major Axis, $a/R_s$    & \nodata   & \multicolumn{2}{c}{\nodata} & \multicolumn{2}{c}{83.2(8)} & \nodata \\
Period, $P$                 & days      & \multicolumn{2}{c}{\nodata} & \multicolumn{2}{c}{32.940045} & \nodata \\
% \hline
Linear LD Coef., $u_1$    & \nodata   & 0.07959 & 0.22941 & 0.09958 & 0.03166  &  0.06363 \\
Quadratic LD Coef., $u_2$ & \nodata   & 0.33761 & 0.39184 & 0.20390 & 0.19793  &  0.11075 \\
\hline
\enddata
\tablecomments{Parentheses indicate $1\sigma$ uncertainties.  The limb darkening (LD) coefficients are held constant using models from the MPS2 grid \citep{Kostogryz2022}. MJD = BJD\sb{TDB} - 2,400,000.5. }
\end{deluxetable*}

We perform a joint fit on the NIRSpec/G395H NRS1 and NRS2 white light curves to derive the orbital parameters used throughout this work. We fit a linear-in-time polynomial and a \texttt{batman} \citep{Kreidberg2015} transit model, allowing the time of transit, inclination, and $a/R_s$ to be jointly constrained by the data. We fix the ExoTiC-LD \citep{Grant2024} limb-darkening coefficients to the values computed from the MPS2 grid \citep{Kostogryz2022}, using stellar parameters of $T_{\mathrm{eff}} = 3457$ K, [M/H] = 0.1, and $\log(g) = 4.8$ dex \citep{Sarkis2018}. \Cref{tab:fit_parameters} lists the best-fit orbital parameters and their uncertainties. For the spectroscopic fits, we hold the orbital parameters and wavelength-dependent limb-darkening coefficients fixed, fitting only for the transit depth and temporal ramp parameters.

\begin{figure*}[t]
    \centering
    \includegraphics[width=1\linewidth]{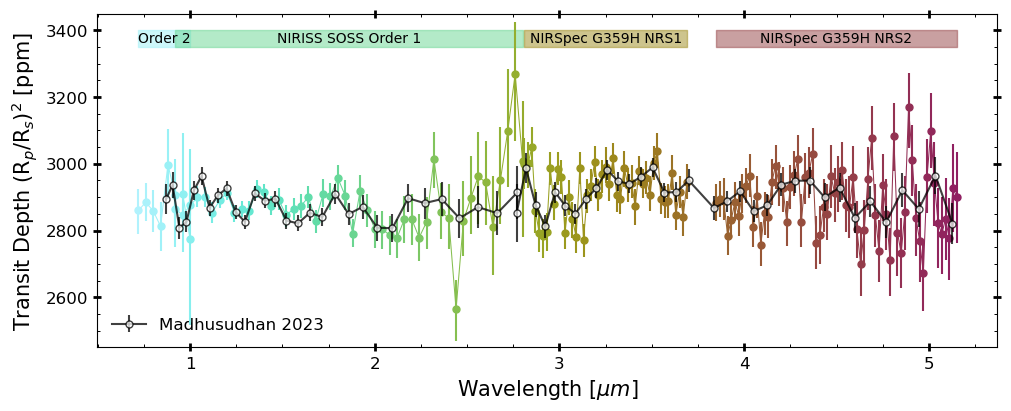}
    \caption{NIRISS/SOSS and NIRSpec/G395H transit spectrum of \planetname.  
    The spectrum matches that published by M23 after applying a -50 ppm offset to their spectrum.}
    \label{fig:nir_spectrum}
\end{figure*}

Following the results of our binning scheme tests for MIRI/LRS, we perform a similar analysis for the NIRSpec/G395H data. While pixel-level odd-even scatter has not been reported in other datasets \citep[e.g.,][]{FuLustig2023,Moran2023,May2023}, we examine the effects of bin size and placement on the resulting NIRSpec transmission spectra for completeness. For both NRS1 and NRS2, we generate light curves using bin widths of 1, 2, 3, 4, 5, 6, 8, 10, 20, 30, and 40 pixel rows. For each bin size, we also explore shifted versions by offsetting the starting pixel. Importantly, we find no dependence of spectral shape or feature presence on bin size or starting location in the NIRSpec/G395H data. For our final spectroscopic fits, we adopt a constant bin width of 20~nm (see \Cref{fig:nir_spectrum}).

\subsection{NIRISS} 
\label{sec:data:niriss} 

We use \eureka v1.2 \citep{Bell2022} to reduce the {\jwst} NIRISS/SOSS data from GO program 2722 (PI: Madhusudhan). This version of \eureka supports \texttt{jwst} pipeline version 1.18.0, and we use CRDS reference context (\textit{pmap}) 1364.

NIRISS/SOSS produces two curved spectral orders on a single detector, with partial overlap. While Stages 1 and 2 of the \texttt{jwst} pipeline do not treat the orders separately, \eureka Stage 3 corrects for the curvature in both traces and extracts time-series stellar spectra for each order. \eureka Stage 4 then generates light curves independently for each order, allowing them to be fit separately. As before, we optimize key reduction parameters in Stages 3 and 4 \citep{Ashtari2025}.

The NIRISS/SOSS light curves exhibit a spot crossing event, so we adopt the orbital parameters derived from the NIRSpec/G395H fit. The spot crossing is within the noise level of the spectroscopic light curves, so we do not attempt to fit the spot itself. We perform the same binning scheme tests for NIRISS/SOSS as we did for NIRSpec/G395H, applying the same bin sizes and shifts to both spectral orders. As with NIRSpec, we find no dependence of spectral shape or feature presence on bin size or starting location for the NIRISS/SOSS data. For our final spectroscopic fits, we adopt a constant bin width of 40~nm. The resulting NIR transit spectrum (\Cref{fig:nir_spectrum}) matches the relative spectrum published by M23; however, there is a 50 ppm offset in the absolute spectrum. The origin of this offset is likely due to differences in the orbital parameters and does not affect our results, as we account for such offsets in our atmospheric retrievals in the next section.

\section{Atmospheric Retrieval Modeling} 
\label{sec:modeling}
%\subsection{Atmospheric Retrievals}

\subsection{Retrieval Setup}

We use the \poseidon \citep{MacDonald2017, MacDonald2023} modeling suite to perform atmospheric retrievals on the planet’s near- and mid-IR transit spectra. Our retrieval framework adopts a common setup that includes free parameters with uniform priors for the following: 
the 10-bar reference radius ($\mathrm{R}_{\mathrm{p, ref}} \sim \mathcal{U}(0.18, 0.24) ~ \mathrm{R_J}$), 
the isothermal atmospheric temperature ($\mathrm{T} \sim \mathcal{U}(100.00, 600.00) ~ \mathrm{K}$), the pressure of an optically thick gray cloud deck ($\log \, \mathrm{P}_{\mathrm{cloud}} \sim \mathcal{U}(-6.00, 2.00) ~ \mathrm{bar}$), and volume mixing ratios ($\log \, \mathrm{X} \sim \mathcal{U}(-12.00, -0.30)$) for 18 molecules \citep[from \poseidon v1.2;][]{Mullens2024}: \ce{H2O}, \ce{CH4}, \ce{CO2}, \ce{CO}, \ce{SO2}, \ce{NH3}, \ce{HCN}, \ce{C2H2}, \ce{C2H4}, \ce{C2H6}, \ce{H2S}, \ce{PH3}, \ce{CS2}, \ce{C2H6S} (DMS), \ce{C6H2S2} (DMDS), \ce{C3H4}, \ce{C2H5Cl}, and \ce{C3H8} \citep{HITRAN-1164, HITRAN2016_XSC, Gordon2017HITRAN2016, HITRAN2016_XSC, HITRAN2020}. This molecular set is motivated by species commonly considered in exoplanet atmospheric retrievals and those of potential relevance to \planetname, based on previous studies \citep[e.g.,][]{Madhusudhan2025, Pica-Ciamarra2025, Welbanks2025, Luque2025}. We added cross sections from HITRAN \citep{HITRAN2016_XSC} for DMS, DMDS, \ce{C3H4}, \ce{C2H5Cl}, and \ce{C3H8} to \poseidon using the \texttt{Cthulhu} Python package \citep{Agrawal2024}.  

We safely omit atmospheric refraction \citep[e.g.,][]{Betremieux2013, Misra2014} in our \poseidon transmission spectrum retrievals. Using the analytic expression from \citet{Robinson2017}, and assuming a H/He-dominated atmosphere, we estimate that refraction could limit sensitivity to pressures deeper than approximately 3 bars for \planetname --- well below the pressures probed in transmission.

We perform a series of retrieval experiments on the available {\jwst} data for \planetname. In the first experiment, we apply the same retrieval setup to eight versions of the MIRI reduction, each with a different starting wavelength corresponding to a shift of 0 -- 7 pixels. In the second experiment, we combine the full 0.7 -- 12~{\micron} {\jwst} transmission spectrum by merging the NIRISS/SOSS, NIRSpec/G395H, and MIRI/LRS datasets, again sampling different MIRI pixel shifts. 
We include free transit depth offsets between detectors ($\delta_{\mathrm{rel}} \sim \mathcal{U}(-1000.00, 1000.00) ~ \mathrm{ppm}$), allowing the NIRSpec NRS1, NRS2, and MIRI/LRS datasets to float relative to the fixed NIRISS/SOSS spectrum. 

\subsection{Retrieval Results}  

\begin{figure*}[t]
    \centering
    \includegraphics[width=0.85\linewidth]{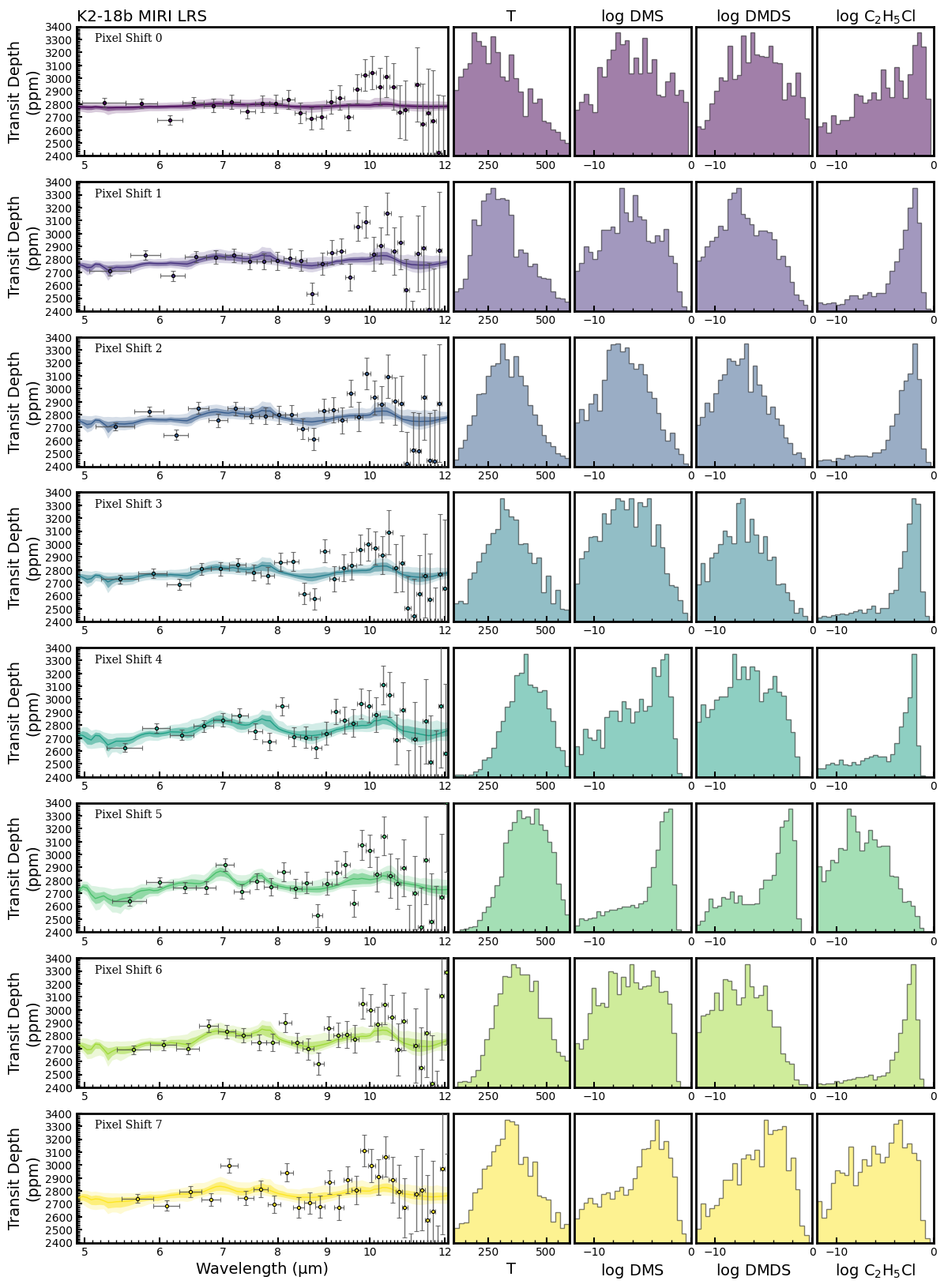}
    \caption{An overview of MIRI retrieval results for different pixel shift cases (rows) using 8 pixel-wide channels. The left column of each row shows a measured transit spectrum along with the median retrieved spectral model within representative $1\sigma$ and $2\sigma$ credibility envelopes. The right subpanels show the inferred 1D marginalized posterior distributions for the isothermal temperature (T), DMS, DMDS, and \ce{C2H5Cl} (chloroethane). Critically, different pixel shift selections lead to different preferences for which gases best fit the spectrum and are constrained, with a trichotomy exhibited here between no abundance constraints (shifts 0,4,7), constraints on DMS+DMDS (shift 5), or constraints on chloroethane (shifts 1,2,3,6). 
    }
    \label{fig:MIRI_retrievals}
\end{figure*}

Retrievals on the \planetname MIRI/LRS spectra alone yield differing trace gas preferences depending on the starting wavelength of the 8-pixel-wide channels. \autoref{fig:MIRI_retrievals} presents the results from our MIRI-only retrievals on the different pixel-shifted spectra described in \autoref{sec:data:MIRI}. These results tend to fall into three broad categories: (1) insignificant posterior constraints on any gases (i.e., flat posteriors; pixel shifts 0, 4, and 7), (2) a slight preference in favor of \ce{C2H5Cl} (pixel shifts 1, 2, 3, and 6), or (3) a slight preference in favor of DMS and/or DMDS (pixel shift 5), as reported by M25. Throughout the text, we define ``favor'' as a quantitative threshold of $\log_{10}(\mathrm{VMR}) > -8$ at $1\sigma$, and ``detect'' as the same threshold at $2\sigma$. While our subsequent retrievals favor several molecules and yield one detection at $2\sigma$, none are significantly detected at $>3\sigma$. The sensitivity of retrieval outcomes to arbitrary choices in the MIRI/LRS data reduction process raises questions about the reliability of atmospheric composition inferences from any single instance. We conclude that past retrieval results from M25 likely contain additional, unquantified uncertainty.

\begin{figure*}[t]
    \centering
    \includegraphics[width=\linewidth]{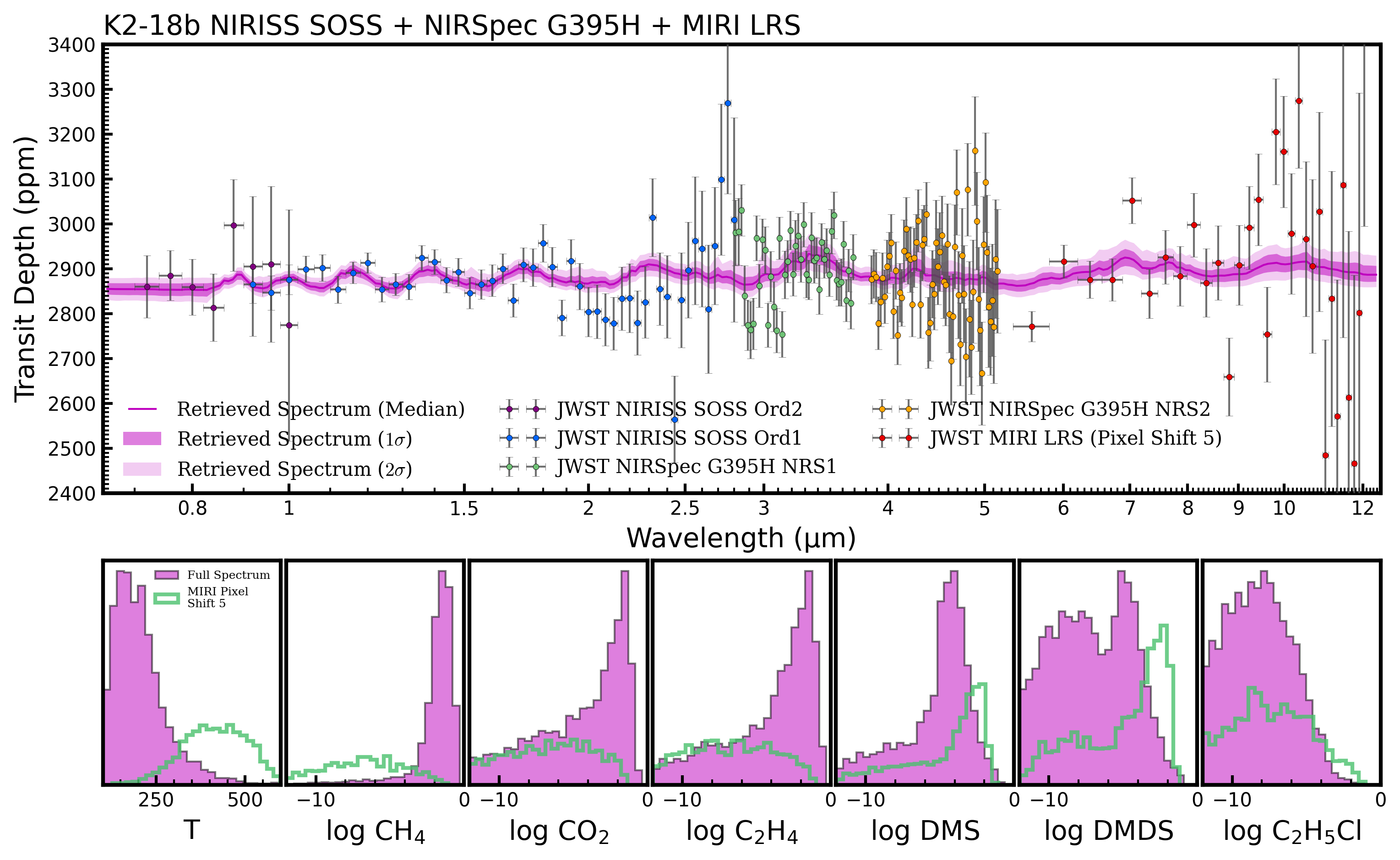}
    \caption{Retrieval results on the full NIRISS/SOSS, NIRSpec/G395H, and MIRI/LRS spectrum (pixel shift 5). The top panel shows the transit spectrum with the median retrieved spectral model and representative $1\sigma$ and $2\sigma$ credibility envelopes. The lower subpanels show the inferred 1D marginalized posterior distributions for the isothermal temperature (T), \ce{CH4}, \ce{CO2}, \ce{C2H4}, DMS, DMDS, and \ce{C2H5Cl}. The complete figure set (pixel shifts 6 and 7) is available in the online journal. \Cref{fig:full_retrievals_6+7} shows additional retrieval results for pixels shifts 6 and 7. 
    }
    \label{fig:full_retrievals_5}
\end{figure*}

\begin{figure*}[t]
    \centering
    \includegraphics[width=\linewidth]{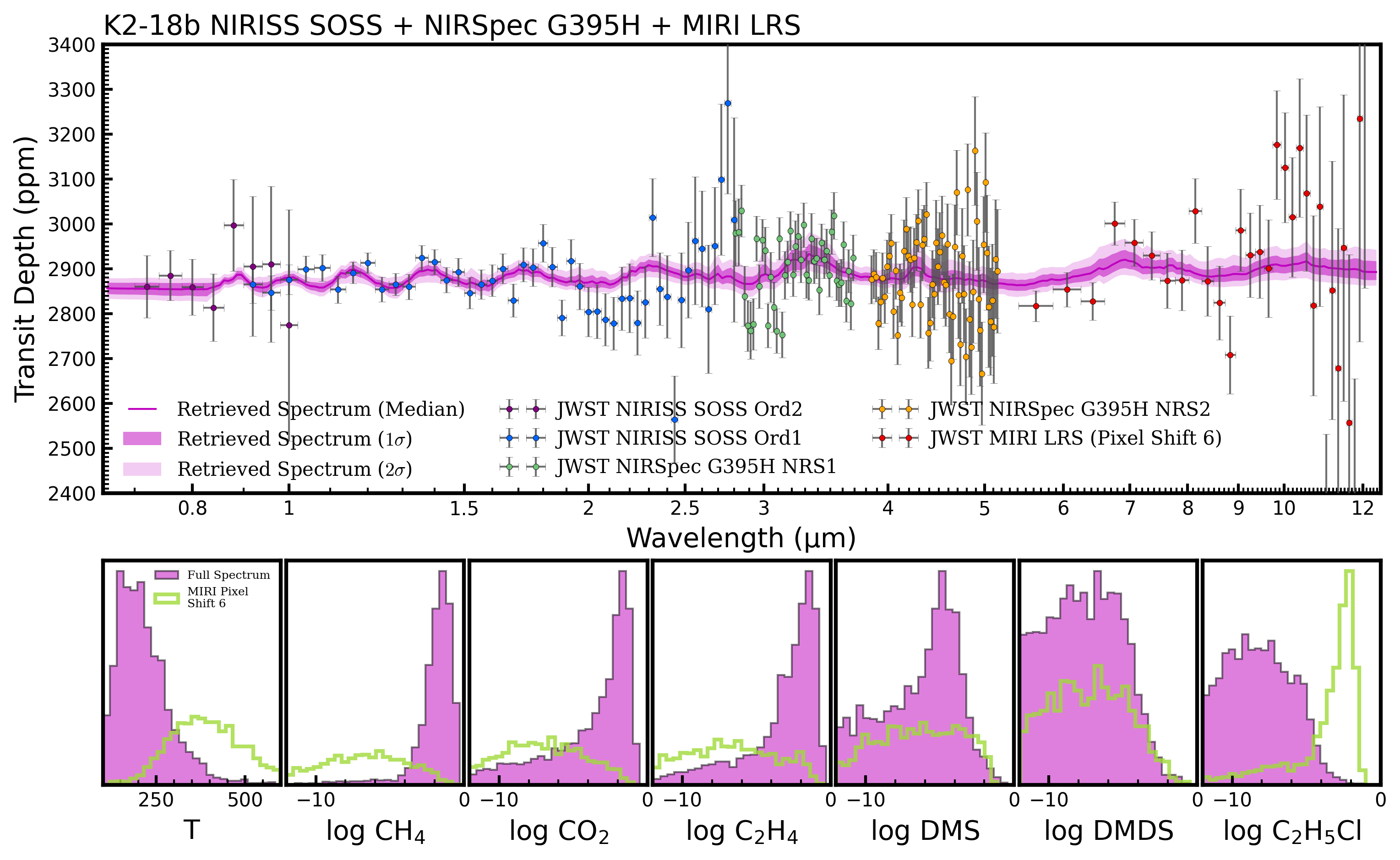}
    \includegraphics[width=\linewidth]{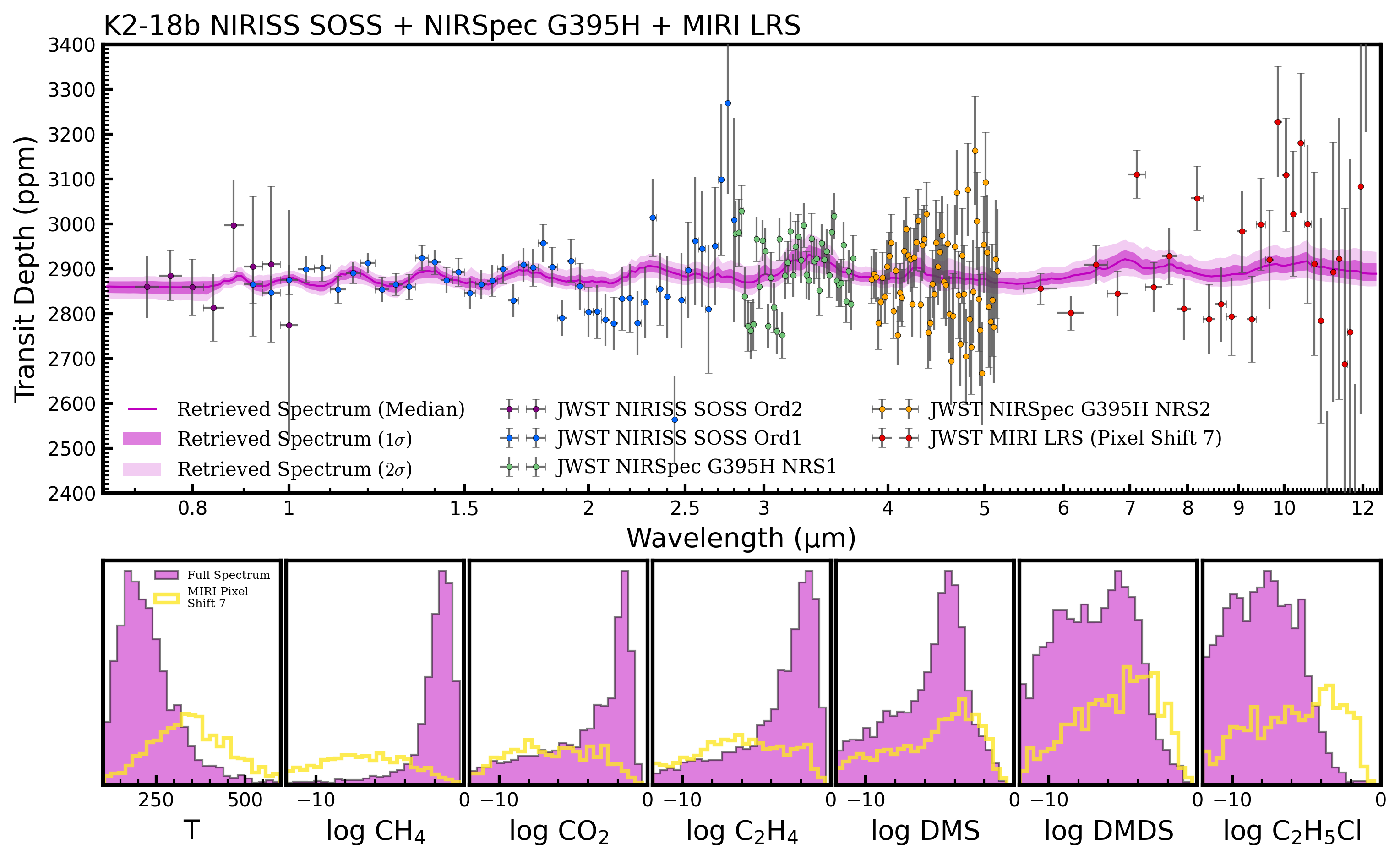}
    \caption{Retrieval results on the full NIRISS/SOSS, NIRSpec/G395H, and MIRI/LRS spectrum (Top: pixel shift 6; Bottom: pixel shift 7). See the caption for \Cref{fig:full_retrievals_5}.
    }
    \label{fig:full_retrievals_6+7}
\end{figure*}

Retrievals on the full 0.7 -- 12~{\micron} \planetname spectrum produce results inconsistent with those from MIRI/LRS data alone, highlighting the significant information gain provided by the shortwave NIR data. \Cref{fig:full_retrievals_5,fig:full_retrievals_6+7} show the results of retrievals using the combined NIRISS/SOSS, NIRSpec/G395H, and MIRI/LRS datasets for pixel shifts 5, 6, and 7. 
% We include three versions as part of a Figure Set, corresponding to MIRI pixel shifts 5, 6, and 7.

In all cases, the MIRI data alone favor hotter isothermal atmospheres, while the inclusion of NIR data yields cooler retrieved temperatures. This temperature discrepancy, consistent with \citet{Luque2025}, is likely caused by systematics in the MIRI/LRS data that introduce spurious features exceeding the true signals, necessitating a hotter temperature to increase the atmospheric scale height.

Our broad-spectrum results detect \ce{CH4} and favor the presence of \ce{CO2}, consistent with M23. We also favor the presence of \ce{C2H4} (ethylene) and/or DMS, though the retrieved abundances of these two species are anti-correlated. This is unsurprising, as both molecules exhibit prominent, overlapping absorption features near 3.3, 7, and 10~\microns. Consequently, \ce{C2H4}, DMS, and a broader suite of hydrocarbons with similar functional groups that absorb at these wavelengths \citep{Luque2025, Niraula2025} remain viable atmospheric candidates for \planetname.

Notably, the MIRI-only retrievals that initially favored \ce{C2H5Cl} (e.g., pixel shift 6) no longer do so when the NIR data are included. Instead, we derive an upper limit on \ce{C2H5Cl} of $-5.69$ ($-4.12$) in $\log_{10}(\mathrm{VMR})$ at $1\sigma$ ($2\sigma$) confidence.

\begin{figure*}[t]
    \centering
    \includegraphics[width=\linewidth]{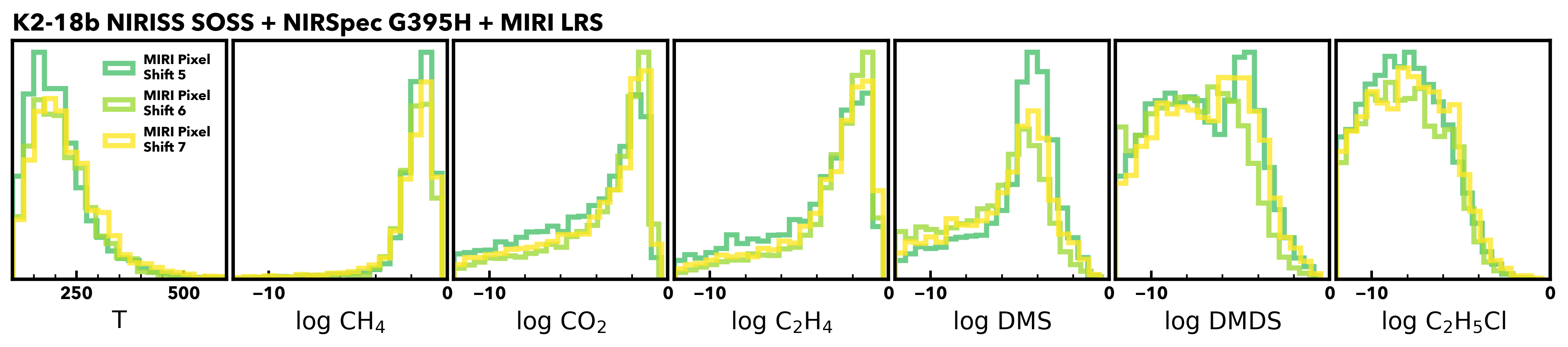}
    \caption{1D marginalized posterior distributions for a subset of retrieval parameters on the full NIRISS/SOSS, NIRSpec/G395H, and MIRI/LRS spectrum for different choices of MIRI pixel shifts. Despite MIRI/LRS pixel shifts resulting in different preferred gases when analyzed individually, these differences are insignificant when combined with NIR data. 
    }
    \label{fig:full_retrieval_hists}
\end{figure*}

\autoref{fig:full_retrieval_hists} compares the 1D marginalized posterior distributions from the three full-spectrum retrievals shown in \autoref{fig:full_retrievals_5}, differing only in the MIRI pixel shift. The resulting posteriors are remarkably consistent, suggesting that the retrieval outcomes are relatively insensitive to the specific MIRI/LRS reduction used due the constraining power of the NIRISS/SOSS and NIRSpec/G395H datasets. 

Based on the available evidence, we conclude that atmospheric inferences for \planetname derived solely on MIRI/LRS data are likely overconfident, and that future observations should be interpreted with caution and in context with the NIR data.

\section{Discussion} 
\label{sec:discussion}

\subsection{Recommendations for MIRI/LRS time-series observations}
\label{sec:miri_sys}

Our reanalysis of the \planetname MIRI/LRS data has uncovered a previously underappreciated systematic effect that likely affects all MIRI time-series observations. Based on the transit depth scatter shown in \Cref{fig:scatter}, it initially appeared that using six or eight pixel rows per channel would sufficiently mitigate instrumental systematics. However, \Cref{fig:MIRI_retrievals} clearly shows that even with eight-pixel-wide bins, different pixel shifts continue to bias the inferred atmospheric composition and temperature, leading to incongruent constraints.
Specifically, four out of the eight pixel shifts favor the presence of chloroethane (\ce{C2H5Cl}), three result in unconstrained posteriors, and only one shift favors the presence of DMS/DMDS. Moreover, the temperature retrieved in the DMS/DMDS case is well-above the planet's equilibrium temperature of 280~K (see \Cref{fig:MIRI_retrievals,fig:full_retrievals_5}).

Given these inconsistencies, we offer the following recommendations for future MIRI/LRS analyses:
\begin{packed_enum}
    \item Use an even number of pixel rows per channel;
    \item Adopt bin sizes larger than eight pixels rows; and
    \item Require that atmospheric retrieval results are both consistent across pixel shifts and physically plausible.
\end{packed_enum}

\subsection{Evaluation of the standards of evidence for \planetname}

In returning to the list of questions in \autoref{sec:intro:evidence}, we now assess whether \planetname{} meets the evidentiary standards required to claim the detection of biosignatures.

For {\bf Q1}, we have shown that the apparent DMS/DMDS signal in the MIRI data is not authentic, but rather an artifact arising from a specific binning scheme. Only 12.5\% of the tested MIRI binning configurations yield support for DMS or DMDS, and inconsistencies in feature amplitudes between the NIR and MIR data further indicate that MIRI-only retrievals are likely fitting noise. These findings are consistent with the lack of strong statistical evidence for spectral features in the MIRI data, as reported by \citet{Taylor2025} and \citet{Welbanks2025}.

For {\bf Q2}, our retrievals show that molecules such as ethylene (\ce{C2H4}) and chloroethane are at least as favored as, and in many cases preferred over, DMS/DMDS, rendering the source of the observed signal ambiguous. This molecular degeneracy aligns with findings by \citet{Luque2025} and \citet{Welbanks2025}, reflecting the large number of plausible chemical species that could produce similar features \citep{Pica-Ciamarra2025, Niraula2025}. Comprehensive follow-up --- including additional {\jwst} transits, improved molecular line lists, and more sophisticated retrieval models --- will be critical to determine whether \planetname truly hosts signs of biological activity or instead exemplifies a biosignature false positive.

For {\bf Q3}, photochemical modeling by \citet{Hu2025} demonstrated plausible abiotic formation pathways for DMS. Hydrocarbons are expected from the photochemical processing of \ce{CH4}, which is robustly detected in \planetname, and in our broad-wavelength retrievals we see hints of \ce{C2H4} and/or DMS. 

A nuance worth highlighting is that biological processes can significantly amplify hydrocarbon production, potentially leading to much higher abundances than abiotic mechanisms alone. In such cases, specific molecular ratios (e.g., \ce{C2H6}/\ce{CH4}) could help discriminate between biotic and abiotic sources \citep{Domagal-Goldman2011}. Unfortunately, the current data lack the precision necessary to constrain such ratios, leaving open the possibility that non-biological processes are responsible for the observed features.

For {\bf Q4}, the inferred DMS abundance of $\sim$100 ppm would require biogenic sulfur production rates that are orders of magnitude higher than those observed on modern Earth \citep{tsai2024}, although DMS remains both chemically and physically plausible in this environment. In contrast, chloroethane --- favored in several MIRI-only retrievals --- is a synthetic compound on Earth, predominantly produced through industrial processes \citep{rossberg2006}. The improbability of its biological production in an exoplanetary context weakens its case as a credible biosignature.

For {\bf Q5}, the MIRI observations alone do not independently detect (or even favor) the presence of DMS. Furthermore, there is no consensus among teams who have reanalyzed the data.  As a result, no independent lines of evidence (through follow-on experiments or analyses from multiple teams) currently support a biological interpretation.

In summary, while the potential for DMS is intriguing --- particularly given its relevance as a potential biosignature gas --- the available data do not meet the evidentiary thresholds defined by the astrobiology community’s five-question framework.

\section{Conclusion}
\label{sec:conclusion}

In this work, we analyzed previously-published, publicly-available {\jwst} observations of \planetname using independent data reduction and spectral retrieval frameworks. Our main findings are as follows:

\begin{itemize}
\item{Independent Reduction and Analysis:}
    Under the same assumptions by M23 and M25, we reproduce the NIRISS, NIRSpec, and MIRI transit spectra they reported.  However, different wavelength binning schemes and pixel shifts in the MIRI data produce a wide variety of apparent planet spectra.  We determine the MIRI spectroscopic features to be unreliable.
\item{Explorations of MIRI Systematics:}
    Our reduction revealed a previously underappreciated systematic in MIRI data that likely affects all time-series observations, not only those of \planetname. We described the nature of this effect and recommended diagnostic tools and mitigation strategies in Section~\ref{sec:miri_sys}. We conclude that the apparent features in the MIRI \planetname spectrum are most likely the result of red noise.
\item{No Evidence of Biosignature Gases:}
    We find no statistically significant evidence for biosignature gases in the atmosphere of \planetname. Moreover, the current data do not satisfy the criteria laid out in the standards of evidence framework (Section~\ref{sec:intro:evidence}) proposed for the robust identification of life.
\end{itemize}

In order to assess the confidence of future biosignature claims, we encourage the exoplanet community to adopt the astrobiology standards of evidence framework and evaluate these claims against all five questions.
 
\begin{acknowledgments}
\textbf{Acknowledgments:}
We would like to thank our anonymous reviewer for their constructive comments and suggestions. This material is based upon work performed as part of the CHAMPs (Consortium on Habitability and Atmospheres of M-dwarf Planets) team, supported by the National Aeronautics and Space Administration (NASA) under Grant No. 80NSSC23K1399 issued through the Interdisciplinary Consortia for Astrobiology Research (ICAR) program. 
This work is based on observations made with the NASA/ESA/CSA James Webb Space Telescope. 
The data were obtained from the Mikulski Archive for Space Telescopes at the Space Telescope Science Institute, which is operated by the Association of Universities for Research in Astronomy, Inc., under NASA contract NAS 5-03127 for JWST. These observations are associated with program \#2722. The specific observations analyzed can be accessed via \dataset[doi:10.17909/gka1-z164]{https://doi.org/10.17909/gka1-z164}.
As a final step before submission, we made use of OpenAI’s ChatGPT to assist with language editing and improving clarity in parts of this manuscript.
\end{acknowledgments}

\facilities{JWST (NIRISS, NIRSpec, MIRI)}

% \texttt{} \citep{};
\software{
\texttt{Astraeus} \citep{astraeus};
\texttt{AstroPy} \citep{astropy2013};
\texttt{BATMAN} \citep{Kreidberg2015};
\texttt{CRDS} \citep{crds};
\texttt{Cthulhu} \citep{Agrawal2024}; 
\texttt{dynesty} \citep{Speagle2020}; 
\texttt{emcee} \citep{Foreman-Mackey2013};
\texttt{ExoTiC-LD} \citep{Grant2024};
\eureka \citep{Bell2022};
\texttt{h5py} \citep{h5py};
\texttt{jwst} \citep{jwst};
\texttt{Matplotlib} \citep{Hunter2007}; 
\texttt{MIRI} \citep{rieke2015};
\texttt{MultiNest} \citep{Feroz2009}; 
\texttt{NumPy} \citep{numpy};
\texttt{Pandas} \citep{reback2020pandas};
\texttt{PySynPhot} \citep{pysynphot}; 
\poseidon \citep{MacDonald2017, MacDonald2023}; 
\texttt{PyMultiNest} \citep{Buchner2014}
\texttt{SciPy} \citep{scipy}
\texttt{Xarray} \citep{hoyer2017xarray}
}.

% \appendix

% \section{Retrieval Extras} 

% \begin{figure*}[t]
%     \centering
%     %\includegraphics[width=0.6\linewidth]{figs/K2-18b_multinst_s5.4_g3_isochem_bigplot_shift5.png}
%     \includegraphics[width=0.85\linewidth]{figs/K2-18b_multinst_s6.4_g3_isochem_bigplot_shift6.png}
%     \includegraphics[width=0.85\linewidth]{figs/K2-18b_multinst_s7.4_g3_isochem_bigplot_shift7.png}
%     \caption{Same as \autoref{fig:full_retrievals}, except for pixel shifts 6 and 7. This plot will be a figure set in the final paper.    
%     }
%     \label{fig:full_retrievals_figset}
% \end{figure*}

\bibliography{main}

\end{document}